\documentclass[twocolumn]{aa}
\usepackage{graphicx}
\usepackage{txfonts}
\begin{document}
\title{Hard X-ray diffuse emission from the Galactic Center seen by
{\it INTEGRAL}.}  \author{A.~Neronov\inst{1,2},
M.~Chernyakova\inst{1,2}, T.J.-L.~Courvoisier\inst{1,2} \and
R.~Walter\inst{1,2} \fnmsep}
         
\institute{INTEGRAL Science Data Center, Ch. d'Ecogia 16, CH-1270
   Versoix, Switzerland \and Geneva Observatory, 51 ch. des
   Maillettes, CH-1290 Sauverny, Switzerland }
 
 \abstract{ We study the hard X-ray (20-100~keV) variability of the
 Galactic Center (GC) and of the nearby sources on the time scale of
 1000~s. We find that 3 of the 6 hard X-ray sources detected by {\it
 INTEGRAL} within the central $1^\circ$ of the Galaxy are not variable
 on this time scale: the GC itself (the source IGR J1745.6-2901) as
 well as the source 1E 1743.1-2843 and the molecular cloud Sgr B2. We
 put an upper limit of $5\times 10^{-12}$~erg/(cm$^2$ sec) (in 20 to
 60 keV band) on the variable emission form the supermassive black
 hole (the source Sgr A*) which powers the activity of the GC
 (although we can not exclude the possibility of rare stronger
 flares). The non-variable 20-100 keV emission from the GC turns out
 to be the high-energy non-thermal tail of the diffuse hard ``8 keV''
 component of emission from Sgr A region. Combining the {\it
 XMM-Newton} and {\it INTEGRAL} data we find that the size of the
 extended hard X-ray emission region is about $20$~pc. The only
 physical mechanism of production of diffuse non-thermal hard X-ray
 flux, which does not contradict the multi-wavelength data on the GC,
 is the synchrotron emission from electrons of energies 10-100~TeV.
  } \date{Received: ; accepted }

\authorrunning{Neronov et al.}

\titlerunning{Hard X-ray diffuse emission from the GC}

\maketitle

\section{Introduction}

Recent observations of the Galactic Center (GC) in 1--10 keV
(\cite{baganoff2001}), 20-100 keV (\cite{belanger2004}),
10~MeV--10~GeV (\cite{mayer-hasselwander1998}) and 1--10~TeV
(\cite{aharonian2004}) reveal the high-energy activity of the nucleus
of the Milky Way.  This activity is powered by a supermassive black
hole (BH) of mass $M_{BH}\simeq 3\times 10^6M_\odot$
(\cite{genzel2000,ghez2000,schoedel2002}) and has a number of puzzling
properties such as an extremely low luminosity, $L_{BH}\simeq
10^{36}$~erg/s (eight orders of magnitude less than the Eddington
luminosity). A number of theoretical models of supermassive BHs
accreting at low rates were put forward to explain the data
(\cite{narayan2002,review,ahaner}) but the ``broad-band'' picture of
activity of the GC is still missing.

Emission from the GC has two major contributions. The first one is the
variable emission from the direct vicinity of the central BH (the
source Sgr A*) detected in X-rays
(\cite{baganoff2001,porquet2003,goldwurm2003}) and in the infrared
(\cite{genzel2003}). The typical variability time scales $T\sim
1$~ksec are close to the light-crossing time of a compact region of
the size of about ten gravitational radii of the central supermassive
BH, $R\sim 10R_{grav}\simeq 10^{13}$~cm.  Apart from the highly
variable emission from a compact object, there is a strong diffuse
component which extends over tens of parsecs around the compact
source.  This diffuse X-ray emission has complex spatial and spectral
properties (\cite{muno2004}). In particular, it contains an extended
``hard'' component which is tentatively explained by the presence of
hot plasma with temperature $T\simeq 8$~keV in the emission region.
However, such an explanation faces serious problems because it is
difficult to find objects which would produce such a hot plasma and,
moreover, this plasma would not be gravitationally bound in the
GC. The origin of the hard component of diffuse X-ray emission can be
clarified using the data on diffuse emission in higher energy bands
(above 10 keV). If the hard diffuse emission is really of thermal
origin, one expects to see a sharp cut-off in the spectrum above 10
keV.

The GC was recently detected in the 20-100 keV energy band by the {\it
INTEGRAL} satellite (\cite{belanger2004}).  The source IGR
J1745.6-2901 is found to be coincident with Sgr~A* to within $\sim
1$~arcmin. However, the wide point spread function of ISGRI imager on
board of {\it INTEGRAL} encircles the whole region of size $\sim
20$~pc around the BH which does not allow to separate the
contributions of Sgr A* itself and of the possible extended emission
around Sgr A*. The only possibility of separation of the two
contributions is to study the time variability of the signal. Indeed,
one expects to see a highly variable signal, if the 20-100 keV
emission from the GC is mostly from the Sgr A*. Moreover, a hint of
the variability in this energy band (a possible 40-min flare) was
reported in (\cite{belanger2004}). At the same time, the data sample
analyzed in (\cite{belanger2004}) was too small to exclude that the
``flare'' is a statistical fluctuation of the signal (the analysis
presented below shows that the latter is the case). The question of
variability of the {\it INTEGRAL} GC source was addressed recently by
(\cite{goldwurm2004}), where the absence of significant variability of
the source at the time scales of 1 day and 1 year was reported and the
previous detection of the ``flare'' by (\cite{belanger2004}) was
attributed to a ``background feature''. It is clear that in order to
find whether the signal detected by {\it INTEGRAL} is from the
supermassive black hole itself or from an extended region around the
GC, it is important to study systematically the variability of the
source at the most relevant time scale of $\sim 1$~ksec (the dynamical
time scale near the black hole horizon). Such a systematic analysis
should allow to separate the ``artificial'' or ``background''
variability which is specific to the coded mask instruments (see
Section \ref{sec:method}) from the true variability of the source.

In what follows we develop a systematic approach to the detection of
variability of {\it INTEGRAL} sources in the ``crowded'' fields, which
contain many sources in the field of view (Section \ref{sec:method}).
We apply the method to the $1^\circ\times 2^\circ$ sky region around
the GC (Section \ref{sec:field}). This region contains six hard X-ray
sources previously detected by {\it INTEGRAL} (\cite{belanger2004}).
We find that the source IGR J1745.6-2901 coincident with the Galactic
Center is not variable in the 20-100 keV energy band. In
Section~\ref{sec:GC} we analyze the spectrum of non-variable hard X-ray
emission from the GC in more details, using the {\it INTEGRAL} and
{\it XMM-Newton} data.  We find that the normalization of the {\it
XMM-Newton} spectrum in 1-10~keV band matches the one of the {\it
INTEGRAL} spectrum only if the {\it XMM-Newton} spectrum is extracted
from an extended region with the radius of $\simeq 6'$ around the GC
position. This indicates that the size of the hard X-ray emission
region is $\sim 10-20$~pc.  In Section \ref{sec:phys} we study the
physical mechanism behind the non-variable hard X-ray emission from
the GC.  Taking into account very special physical conditions in the
GC region, we show that the only viable mechanism is the synchrotron
emission from electrons of energies of 10-100~TeV. We find that
inverse Compton and bremsstrahlung emission from such electrons are at
the level of the detected TeV flux from the GC.

\section{Method for the detection of variability}
\label{sec:method}

In principle, the variability of {\it INTEGRAL} sources can be
analyzed in a standard way studying (in)consistency of the detected
signal with the one expected from a non-variable source.  The {\it
INTEGRAL} data are naturally organized by pointings of duration of
$\sim 1-3$~ksec in average (Science Windows, ScW). The simplest way to
detect the variability of a source on the ksec time scale is therefore
to analyze the evolution of the flux from the source on ScW-by-ScW
basis.  If the flux from a source in the $i$-th ScW is $F_i$ and the
flux variance is $\sigma_i$, the $\chi^2$ of the fit by a constant is
$\chi^2=\sum_{i=1}^N(F_i-\overline F)^2/\sigma^2_i$ (here $\overline
F=\left.(\sum_{i=1}^N F_i/\sigma_i^2)\right/(\sum_{i=1}^N
1/\sigma_i^2)$ is the weighted average flux and $N$ is the total
number of ScWs).  The probability of obtaining a given $\chi^2$ under
the assumption that the source is non-variable is ${\cal
P}=1-P(N-1,\chi^2)$ (here $P(a,x)$ is the incomplete gamma-function).

However, in reality, if one applies the above method to the analysis
of of ``crowded'' fields which contain many sources, one would obtain
a surprising (wrong) result that {\it all} the detected sources are
highly variable! The reason for this is that {\it INTEGRAL} is a coded
mask instrument. In the coded mask instruments each source casts a
shadow of the mask on the detector plane. Knowing the position of the
shadow one can reconstruct the position of the source on the sky.  If
there are several sources in the field of view, each of them produces
a shadow which is spread over the whole detector plane. Some detector
pixels are illuminated by more than one source. If the signal in a
detector pixel is variable, one can tell only with certain
probability, which of the sources illuminating this pixel is
responsible for the variable signal. Thus, in a coded mask instrument,
the presence of bright variable sources in the field of view
introduces artificial variability for almost all the other sources in
the field of view. Moreover, since the overlap between the shadowgrams
of the bright variable source and of the sources at different
positions on the sky varies with the position on the sky, one can not
know ``in advance'' what is the level of ``artificial'' variability in
a given region of the deconvolved sky image.

In order to overcome this difficulty, one has to measure the
variability of the flux not only directly in the sky pixels at the
position of the source of interest, but also in the background pixels
around the source. Obviously, the ``artificial'' variability
introduced by the nearby bright sources is the same in the background
pixels and in the pixel(s) at the source position. Practically, one
can calculate the $\chi^2$ for each pixel of the sky image and compare
the values of $\chi^2$ at the position of the source of interest to
the mean values of $\chi^2$ in the adjacent background pixels. The
variable sources should be visible as local excesses in the $\chi^2$
map of the region of interest.

If a source can be localized in the variability image one can estimate
the ratio between the average flux, $\overline F$ and the amplitude of
the flux variations, $\Delta F$.  One can assume, for the sake of the
argument, that the variations are normally distributed around the
average value with the typical variance $\Sigma_s^2$. Assuming that
the variations of the background and of the source flux are
independent one can ascribe the excess scatter of the flux data points
at the location of the source to the source variability. If the
typical variance of the background in each ScW is $\Sigma_b^2$, the
resulting variance in the image pixel containing the source is
$\Sigma^2$(Ra$_s$,~Dec$_s$)$=\chi^2_{red}$(Ra$_s$,~Dec$_s)\Sigma_b^2/N=(\Sigma_b^2+\Sigma^2_s)/N$. One
can estimate the source variability as $V=\Delta F/\overline
F=\sqrt{(\chi^2_{red}(Ra_s,Dec_s)-1)\Sigma_b^2}/\overline F$. Note,
that if a source is not detected in the variability image, this means
that typical variations of its flux are $\Delta F\le
\sqrt{\Delta\chi^2\Sigma_b^2}$ where $\Delta\chi^2$ is the typical
scatter of the values of $\chi^2$ for the background pixels.  Taking
into account that for the large number of ScWs (large $N$)
$\Delta\chi^2\sim N^{-1/2}$, one can estimate the variable
contribution to the flux as $\Delta F\le \sqrt{\Sigma_b^2/N}$.

Note that the above upper limit on the variable contribution to the
flux is of the order of the square root of the variance of the mean
value $\sqrt{\Sigma^2}$. It is known that with the current version of
OSA (4.2), the systematic effects start to contribute significantly to
the variance when the exposure time is more than $10^5$ s
(\cite{arash}). This leads to a slower than $N^{-1/2}$ decrease of the
variance with the increasing number of ScWs. The same should apply for
the variability upper limit $\Delta F$.

\vskip0.5cm
\section{Variability of the sources in GC region}
\label{sec:field}

\begin{figure}
\resizebox{\hsize}{!}{\includegraphics[angle=0]{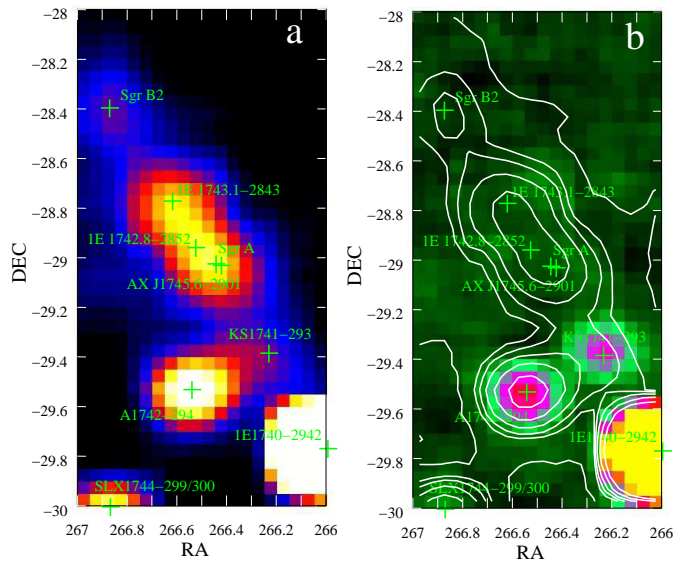}}
\caption{ (a). Intensity image of the GC region in 20-60 keV energy
band. Black corresponds to zero flux, white to the flux $\ge 1$
cnt/s. Total exposure time is about $2.4$ Ms. (b)20-60 keV variability
image of the GC region. Value of each pixel corresponds to the reduced
$\chi^2$ value for the fit of the detected flux in this pixel by a
constant. Black corresponds to the values of $\chi^2=1$ while yellow
corresponds to $\chi^2=4$. Contours show the 20-60 keV flux (in
logarithmic scale) from the same region of the sky. }
\label{fig:20-60}
\end{figure}

\noindent 
We have applied the above method to the $1^\circ\times 2^\circ$ sky
region centered on the GC. Using the publicly available data (see
{\it http://isdc.unige.ch}) with effective exposure time at the
position of the GC of more than $1$~Msec we have produced the hard
X-ray images of the region of interest with the Offline Software
Analysis (OSA) package (\cite{courvoisier2004}), version
4.2. Fig.~\ref{fig:20-60}a shows the 20-60~keV intensity map of the
field around the GC. The size of each pixel in the images is $3'\times
3'$. The value of each pixel of the image is the weighted average
$\overline F$ of the flux from the sky direction corresponding to the
center of the pixel. Fig.~\ref{fig:20-60}b is the $\chi^2_{red}$
(variability) map of the same region in the same energy band.

One can see that from all sources visible on the intensity map,
Fig.~\ref{fig:20-60}a, only 3 appear at the variability map,
Fig.~\ref{fig:20-60}b. Among the variable sources are the well-known
X-ray binaries, 1E 1740.7-2942, A 1742-294 and KS 1741-293. The
variability indexes $V=\Delta F/F$ for these sources are $0.07$,
$0.22$ and $0.59$, respectively. One can see that in spite of large
excess $\chi^2_{red}$ at the location of the brightest source, the
X-ray binary 1E 1740.7-2942, the source flux varies at the level of
only about 7\%. This implies an important limitation for the possible
applications of the method of variability search described above: the
X-ray binaries which produce much lower flux (e.g. at the level of KS
1741-293) but have similar variability properties as 1E 1740.7-2942,
would not be localized in the variability map, because the excess of
$\chi^2_{red}$ produced by such sources would be less than the
statistical scatter of the $\chi^2_{red}$ values over the image
pixels.

In this respect the source 1E 1743.1-2843 is an interesting example.
As one can see, this source is not detected in the variability image.
Calculating the upper limit on the variable fraction of the source
flux, $\Delta F$, along the lines explained above, one gets an upper
limit on possible variability at the level of $8$\%. 1E 1743.1-2843
was discovered by the Einstein Observatory (\cite{watson}). Its
spectral characteristics suggest an interpretation as a neutron star
LMXB (\cite{cremonesi1999}) while its luminosity suggests that the
source should belong to the Atoll-class sources and should exhibit
Type-I X-ray bursts. The lack of substantial variations of the X-ray
flux from this source during the last 20 yr of observations questions
such an interpretation (\cite{porquet2003a}). The lack of variability
in 20-60 keV energy band also argues against the neutron-star X-ray
binary interpretation for this source, although a definite conclusion
would be possible only after a systematic study of variability
properties of LMXBs detected by the {\it INTEGRAL}.

\vskip0.5cm
\section{Non-variability of the GC source.}
\label{sec:GC}

The GC itself (the source IGR J1745.6-2901) also does not appear in
the variability map. The upper limit on the variable contribution to
the flux can be obtained, as it is explained above, from the typical
value of the variance in a single ScW of duration of $\sim 2$~ksec in
20-60~keV band (\cite{arash}), $\Sigma_b^2\simeq 10^{-10}$~erg/(cm$^2$
s). Dividing $\Sigma_b^2$ by $\sqrt{N}$ and taking into account the
systematic ``renormalization'' by a factor $\simeq 1.4$ which has to
be applied to the image Fig.~\ref{fig:20-60}b (see the discussion at
the end of the previous section), we find an upper limit to the
variable contribution at the level of $V=\Delta F/F\le 0.08$, or,
equivalently, $\Delta F\le 5\times 10^{-12}$~erg/(cm$^2$~sec) in the
20 to 60 keV band.

This upper limit is obtained under certain assumption about the
variability type (Gaussian fluctuations around the mean flux). At the
same time, if the variable emission is of different type, the above
upper limit would not apply.  For example, if the source has exhibited
one or two powerful flares during the whole observation period, the
amplitude of the flares can not be deduced from the statistical
fluctuations of the signal.

The non-detection of variability of IGR J1745.6-2901 indicates that
the main contribution to the hard X-ray flux from this source possibly
comes from an extended region around the supermassive BH, rather than
from the direct vicinity of the BH horizon. Indeed, simple powerlaw
extrapolation of the observed X-ray flux from Sgr A* shows that during
the brightest flares the 10-100~keV flux would be roughly at the level
detected by the {\it INTEGRAL}. However, in this case the hard X-ray
flux should vary by $\sim 100$\% at the ksec time scale, which is well
beyond the upper limit found above. At the same time, the flux in 1-10
keV energy band is dominated by the diffuse emission and a simple
powerlaw extrapolation of the diffuse X-ray signal to hard X-ray band
is in good agreement with the {\it INTEGRAL} data (see below).

As it is discussed in the Introduction, the diffuse X-ray emission
from the region around Sgr A* contains a hard component of unclear
nature. The 8~keV ``temperature'' of this hard component is at the
upper end of the energy band probed by the X-ray telescopes and in
order to prove the thermal nature of this component one has to observe
the exponential cut-off in the spectrum above 10~keV.  The analysis of
{\it {\it INTEGRAL}} spectrum of IGR J1745.6-2901 shows that such a
sharp cut-off is not observed.

\begin{figure}
\begin{center}
\resizebox{0.9\hsize}{!}{\includegraphics[angle=0]{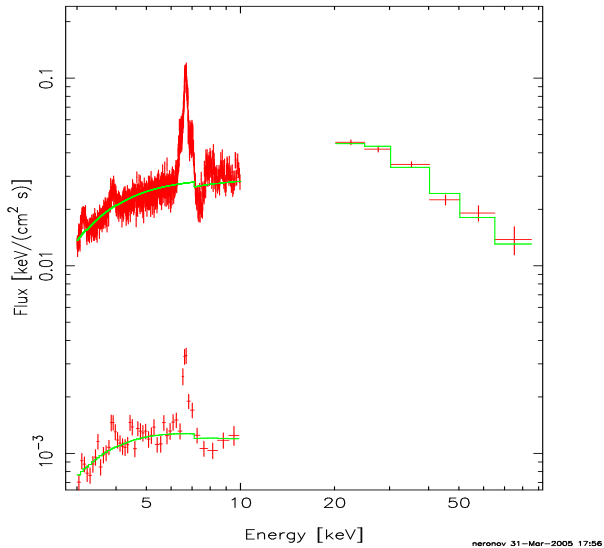}}
\caption{ ISGRI spectrum of IGR J1745.6-2901 (the GC) in the 20-200
keV energy band. Also shown is the {\it XMM-Newton} spectrum (diffuse
emission plus point sources) from the circular region of the radius
$6'$ (about the size of the point spread function of ISGRI). Strong
line in the {\it XMM-Newton} data is the 6.4~keV Fe K$_{\alpha}$
line. For comparison we show also the {\it XMM-Newton} spectrum
extracted from a circular region of the radius 15'' centered on Sgr A*
(the lower curve in 3-10 keV band).}
\label{fig:sgr_spectrum}
\end{center}
\end{figure}

The 20-200~keV spectrum of IGR J1745.6-2901 is shown on
Fig.~\ref{fig:sgr_spectrum} together with {\it XMM-Newton} spectrum in
3-10~keV energy band extracted from a circular region of radius $R=6'$
around the GC (chosen to match the angular resolution of ISGRI imager
on board of {\it {\it INTEGRAL}}).
The {\it XMM-Newton} Observation Data Files (ODFs), {\sc
obs\_id}~0111350101, were obtained from the online Science Archive
{\it
http://xmm.vilspa.esa.es/external/xmm\_data\_acc/xsa/index.shtml}.
The data were processed and the event-lists filtered using {\sc
xmmselect} within the Science Analysis Software ({\sc sas}) v6.0.1.
The 20-200~keV spectrum can be fit by a simple power law model with
the photon spectral index $\Gamma_{20-200}=3.0\pm 0.1$ and flux
$F_{20-200}=(4\pm 2)\times 10^{-11}$~erg/(cm$^2$ s). However, a broken
power-law provides a better fit to the {\it {\it INTEGRAL}} data. The
broken power law with a break energy at $E_{break}=26\pm 1$~keV is
characterized by a low energy spectral index
$\Gamma_{low}=1.85^{+0.02}_{-0.06}$, and a high-energy spectral index
$\Gamma_{high}=3.3\pm 0.1$. It provides the best simultaneous fit to
the {\it XMM-Newton} and {\it {\it INTEGRAL}} data. The
intercalibration factor between the two instruments is just $1.2$
which indicates that 20-200~keV flux detected by {\it {\it INTEGRAL}}
matches well the 1-10~keV flux collected from the region of the size
about the point spread function of ISGRI.
\begin{figure}
\begin{center}
\resizebox{0.9\hsize}{!}{\includegraphics[angle=0]{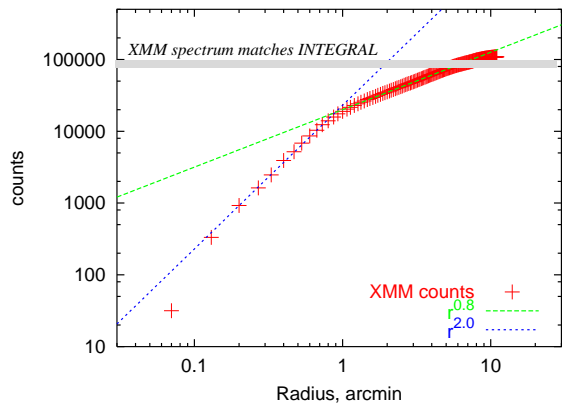}}
\caption{Dependence of the (background corrected) X-ray flux collected
  from a disk on the radius of the disk centered at Sgr A* in the {\it
  XMM-Newton} image of the Galactic Center. Thin solid line shows the
  fit $F\sim r^{0.8}$ for the radius range $1'<r<10'$. Dashed line
  shows the fit $F\sim r^{2}$ for the radius range $r<1'$. Thick gray
  line shows the flux level at which the {\it XMM-Newton} spectrum
  matches the {\it INTEGRAL} spectrum with inter-calibration factor
  $\simeq 1$. }
\label{fig:radial}
\end{center}
\end{figure}

It is important to note that because of the presence of diffuse
emission component around the Galactic Center, the normalization of
the {\it XMM-Newton} flux depends strongly on the size of the region
from which the spectrum is extracted. For example, if one collects the
flux from a disk-like region of the radius $r$ centered on the
Galactic Center, the flux grows as $r^2$ for $r<1'$, see
Fig.~\ref{fig:radial}. From Fig.~\ref{fig:radial} one can see that the
mismatch between the normalization of the {\it INTEGRAL} spectrum and
{\it XMM-Newton} spectrum collected from the $r\simeq 1'$ disk would
be a factor of $\simeq 10$. For the disk radii $r>1'$ the flux grows
proportionally to $r$. The fact that the {\it XMM-Newton} spectrum
matches the {\it INTEGRAL} spectrum when the disk radius reaches
$r\simeq 6'$ has an important implication for the physics of diffuse
hard X-ray emission.

\section{The nature of extended non-thermal hard X-ray emission around
  the GC.}
\label{sec:phys}
\begin{figure}
\resizebox{\hsize}{!}{\includegraphics[angle=0]{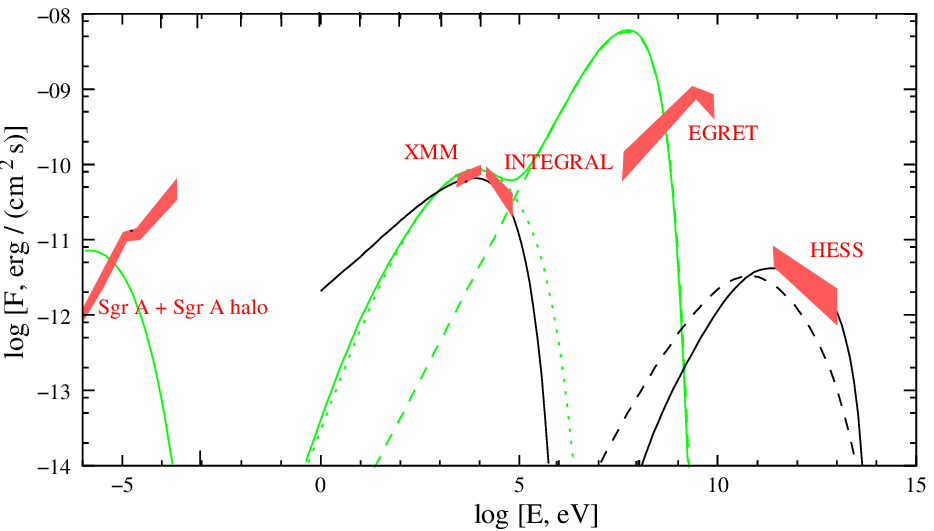}}
\caption{Broad band spectrum of diffuse emission from the Galactic
  Center region. Radio data are from (\cite{pedlar1989}), HESS data
  are from (\cite{aharonian2004}) and EGRET data are from
  (\cite{mayer-hasselwander1998}).  Thin (black) line shows
  synchrotron and IC emission from electron powerlaw electron
  distributions with cutoff energy $E_0=30$~TeV, in magnetic field
  $B=3\times 10^{-4}$~G. Thin dashed line is the bremsstrahlung
  emission from the same population of electrons, assuming typical
  matter density $n\simeq 10^4$~cm$^{-3}$.  Thick (green) line: the
  cut-off energy $E_0= 3$~GeV, magnetic field $B=3\times
  10^{-5}$~G. Thick dashed line shows the bremsstrahlung component
  assuming the density $n\simeq 10^3$~cm$^{-3}$.  }
\label{fig:s-IC-b}
\end{figure}

We have seen in the previous sections that several complementary
arguments indicate that the hard X-ray emission from the GC detected
by {\it INTEGRAL} originates from an extended region around the
supermassive black hole. The inter-calibration factor of order of $1$
between {\it XMM-Newton} and {\it INTEGRAL} spectra can be achieved
only if the {\it XMM-Newton} flux is collected from a circular region
of the radius $r\simeq 6'$. This means that the size of the hard X-ray
extended emission region is about $D\simeq 20$~pc.

In order to understand the mechanism of diffuse nonthermal 10-100 keV
emission from the inner $\sim 10$~pc of the Galaxy, it is useful to
recall that this region is characterized by quite special
astrophysical conditions. It is quite densely populated with giant
molecular clouds and the typical gas/dust density throughout the
region is $n\sim 10^4$~cm$^{-3}$ (see (\cite{metzger1996}) for a
review). Most of the estimates of magnetic field are in the range
$B\sim 10^{-4}-10^{-3}$~G (\cite{yusef-zadeh1996}), much stronger than
the typical Galactic magnetic field.  Besides, the density of the
infrared-optical background in this region, $U_{ir-o}\sim
10^2-10^3$~cm$^{-3}$ (\cite{cox1988}), is more than two orders of
magnitude higher than the density of cosmic microwave background
radiation. Such ``exotic'' conditions should be taken into account in
the modeling of physical processes leading to the emission of diffuse
non-thermal X-rays.

Among the possible emission mechanisms, synchrotron radiation from
electrons of energies $E_e\sim 10^{13}-10^{14}$~eV is the most
plausible candidate. The typical energy of synchrotron photons
produced by such electrons is $E_{syn}\simeq
50(B/10^{-4}$~G$)(E_e/10^{14}$~eV$)^2$~keV. The synchrotron cooling
time of electrons emitting at 10~keV is $t_{syn}\simeq 16
(B/10^{-4}$~G$)^{-3/2}(\epsilon/10$~keV$)^{-1/2}$~yr.  One can see
that the cooling time is too short for electrons accelerated near the
supermassive black hole to spread over the emission region of the size
$D\sim 20$~pc if one assumes diffusion of electrons injected by a
central source. Naively, to overcome this difficulty, one would assume
a lower magnetic field strength in the emission region. However, in
this case the inverse Compton (IC) flux from electrons which emit
synchrotron radiation at $\sim 10$~keV would be stronger than the TeV
flux from the GC detected by HESS. Thus, within the synchrotron
scenario one has to assume that either magnetic field in the emission
region is mostly ordered, so that electrons escape along the magnetic
field lines, rather than diffuse in the random magnetic field, or that
10-100 TeV electrons are injected not by a point source at the
location of the GC, but throughout the whole extended region of the
size $\sim 10-20$~pc. Possible mechanism leading for such ``extended
injection'' is e.g. cascading of the 1-100~TeV gamma quanta on dense
infrared photon background in the central 20~pc of the Galaxy
(\cite{neronov2002,ahaner}). Otherwise, electrons can be accelerated
in the shell of supernova remnant Sgr A East, whose size is about
5~pc.

 Large density of the infrared-optical photon background and of the
molecular gas in the emission region lead to significant IC and
bremsstrahlung emission from the $10^{13}-10^{14}$~eV electrons. The
results of calculation of the broad-band spectrum
synchrotron-IC-bremsstrahlung emission from the inner 20~pc of the
Galaxy are shown in Fig.~\ref{fig:s-IC-b}. One can see that the
expected IC and bremsstrahlung fluxes are strong enough to match the
observed level of the TeV emission from the GC. If the TeV flux
detected by HESS is produced via the above mechanism, it should be not
variable, since the IC and bremsstrahlung flux also come from the
extended region of the size of several tens of kiloparsecs. The
(non)variability of the TeV signal from the GC can be tested with
future HESS observations.

Since the synchrotron emission results in efficient cooling of the
high-energy electrons, all the power supplied by the supermassive
black hole is dissipated with almost 100\% efficiency in the form of
the hard non-thermal X-ray emission. The required power of the
supermassive black hole in the ``synchrotron-IC-bremsstrahlung''
scenario is just $P_{syn}\simeq 10^{36}$~erg/s, about the total power
of Sgr A* observed in infrared. Thus, the above scenario is the most
``economic'' one.

Other possible mechanisms of non-thermal 10-100~keV emission appear to
be much less efficient. For example, if one tries to explain the
observed hard X-ray flux with the bremsstrahlung, one finds
immediately that for moderately relativistic electrons which can emit
bremsstrahlung radiation at $20-50$~keV, the bremsstrahlung cooling
rate is some 5 orders of magnitude less than the Coulomb loss rate,
which means that the bremsstrahlung is very energetically inefficient
mechanism of powering the nonthermal hard X-ray emission. The
supermassive black hole should produce the power at the level of
$L_{BH}\ge 10^5L_{X-ray}\sim 10^{41}$~erg/s in this scenario. Although
such a luminosity is still much below the Eddington luminosity of a
$4\times 10^6M_\odot$ black hole, it is orders of magnitude higher
than the observed luminosity of Sgr A* in the wide photon energy range
from radio to very-high-energy gamma-rays.

The possibility that the observed non-thermal emission is produced via
IC scattering of the dense infrared photon background is also
unsatisfactory. The IC cooling time for electrons emitting in the
10-100~keV band, $t_{IC}\simeq 3\times 10^6\left(U_{ir-o}/(100\mbox{
eV/cm}^3)\right)^{-1}(E_e/1$~GeV$)^{-1}$~yr, is much larger than the
bremsstrahlung cooling time, $t_{br}\simeq 4\times
10^3(n/10^4$~cm$^{-3})^{-1}$~yr. This means that (1) the supplied
power should be 2-3 orders of magnitude higher than the observed hard
X-ray luminosity (depending on the assumptions about the dust density)
and (2) the hard X-ray IC emission should be accompanied by much
stronger 100 MeV bremsstrahlung emission. In fact, the predicted
bremsstrahlung flux is at least an order of magnitude higher than the
GC flux found in the 100~MeV-GeV band by EGRET (see
Fig.\ref{fig:s-IC-b}).

Thus, the synchrotron radiation is the most probable mechanism of the
diffuse hard X-ray emission from the inner 20~pc of the Galaxy. In
order to test the above synchrotron-IC-bremsstrahlung scenario, one
has to study the correlation of the spatial distributions of
20-100~keV and 1-10 TeV fluxes.  Although it is quite difficult to
analyze extended sources with the coded mask instruments, like {\it
INTEGRAL}, it would be interesting to extract the information on the
extension of the source directly from the 10-100~keV data (not from
the matching with the lower-energy spectrum, like it is done
above). We leave this question for future study.

\end{document}